\documentclass{mem}
\usepackage{natbib}\usepackage{txfonts}\usepackage{balance}
\usepackage{graphicx}
\usepackage[a4paper,breaklinks]{hyperref}
\idline{00}{000}
\begin{document}
\def\teff{$T\rm_{eff }$}
\def\kms{$\mathrm {km s}^{-1}$}

\title{
The X-Gamma Imaging Spectrometer (XGIS) \\ onboard THESEUS 
}

   \subtitle{}

\author{
R.~Campana\inst{1} 			\and
F.~Fuschino\inst{1} 		\and
C.~Labanti\inst{1}			\and
L.~Amati\inst{1}			\and
S.~Mereghetti\inst{2}		\and
M.~Fiorini\inst{2}			\and
F.~Frontera\inst{1,3}		\and
G.~Baldazzi\inst{4}			\and
P.~Bellutti\inst{5}			\and
G.~Borghi\inst{5}			\and
I.~Elmi\inst{6}				\and
Y.~Evangelista\inst{7}		\and
M.~Feroci\inst{7}			\and
F.~Ficorella\inst{5}		\and
M.~Orlandini\inst{1}		\and
A.~Picciotto\inst{5}		\and
M.~Marisaldi\inst{1,8}		\and
A.~Rachevski\inst{9}		\and
M.~Uslenghi\inst{2}			\and
A.~Vacchi\inst{9}			\and
G.~Zampa\inst{9}			\and
N.~Zampa\inst{9}			\and
N.~Zorzi\inst{5}			
}

\institute{
INAF/IASF-Bologna, Via Gobetti 101, I-40129, Bologna, Italy
\and
INAF/IASF-Milano, Via Bassini 23, I-20100, Milano, Italy
\and
Universit\`a di Ferrara, Dip. di Fisica, via Saragat 1, I-44122, Ferrara, Italy
\and
Universit\`a di Bologna, Dip. di Fisica, viale Berti Pichat 6, I-40127, Bologna, Italy
\and
Fondazione Bruno Kessler, via Sommarive 8, I-38123, Trento, Italy
\and
CNR/IMM, Via Gobetti 101, I-40129, Bologna, Italy
\and
INAF/IAPS, Via Fosso del Cavaliere 100, I-00133, Roma, Italy
\and
University of Bergen, Bergen, Norway
\and
INFN -- Sezione di Trieste, Padriciano 99, I-34127, Trieste, Italy
\\
\email{campana@iasfbo.inaf.it}
}

\authorrunning{Campana et al.}
\titlerunning{The XGIS instrument onboard THESEUS}

\abstract{
A compact and modular X and gamma-ray imaging spectrometer (XGIS) has been designed as one of the instruments foreseen onboard the THESEUS mission proposed in response to the ESA M5 call. 
The experiment envisages the use of CsI scintillator bars read out at both ends by single-cell 25 mm$^2$ Silicon Drift Detectors. 
Events absorbed in the Silicon layer (lower energy X rays) and events absorbed in the scintillator crystal (higher energy X rays and $\gamma$-rays)
are discriminated using the on-board electronics. A coded mask provides imaging capabilities at low energies, thus allowing a compact and sensitive instrument in a wide energy band ($\sim$2 keV up to $\sim$20 MeV).
The instrument design, expected performance and the characterization performed on a series of laboratory prototypes are discussed.
}
\maketitle{}

\section{Introduction} 
Gamma-Ray Bursts (GRBs) are one of the most intriguing and challenging phenomena for modern science. 
Because of their huge luminosities up to more than 10$^{52}$ erg/s, 
their redshift distribution extending from $z\sim0.01$ up to $z>9$ (i.e., much above that of supernovae of the Ia class, SNe Ia, and galaxy clusters), 
their association with peculiar core-collapse supernovae and with neutron star/black hole mergers, 
their study is of very high interest for several fields of astrophysics. 
These include, e.g., the physics of matter in extreme conditions and plasma physics, black hole physics, core-collapse SNe, cosmology and fundamental physics, production of gravitational wave signals. 
Despite the huge observational advances occurred in the last twenty years, several open issues still affect our comprehension of these phenomena, and their exploitation for fundamental physics and cosmology. 
Among the most relevant aspects are all those connected with the so called ``prompt'' emission. 
A better knowledge of the involved emission processes and of the source geometry is essential, not only to clarify the origin of the ``central engine'' and its connection with the progenitors, but also to assess the real energy budget of different classes of GRBs eventually allowing us to use the GRBs for fundamental physics and cosmology studies. 

To address these fundamental issues, time resolved spectroscopy (and possibly polarimetry) of the GRB prompt emission over a broad energy range from $\sim$1--2 keV (i.e. well below the range of past, present and near future GRB detectors) to several MeVs is needed.

In the framework of the THESEUS mission, we are therefore working towards the design of a monolithic system (\emph{X and Gamma Imaging Spectrometer}, XGIS) which would allow detection, localisation, spectroscopy and timing of GRBs and other high energy transients, like soft gamma-ray repeaters, over an unprecedented broad energy band.
The goal is to obtain a detector with unique capabilities, such as the combination of a low energy threshold (1--2 keV) and energy resolution significantly better than that of any other GRB detection system based on scintillators \citep[e.g., Fermi/GBM,][]{gbmpaper} or CZT/CdTe semiconductors (INTEGRAL/ISGRI, \citealt{isgripaper}; Swift/BAT, \citealt{batpaper}), besides timing capabilities down to a few micro-second resolution over the whole energy band, and possible polarimetric and Compton telescope capabilities thanks to the true three-dimensional photon interaction reconstruction.

In this paper we describe the design of such an instrument. In Section~\ref{s:theseus} the science case of the THESEUS mission is briefly outlined, while in Section~\ref{s:xgis_arch} we describe both the XGIS design and the laboratory work on several prototypes, and in Section~\ref{s:performance} we discuss the instrument performance, while in Section~\ref{s:conclusions} we draw our conclusions.

\section{The THESEUS mission}\label{s:theseus}
\begin{figure*}[htbp]
\centering
\includegraphics[width=\textwidth]{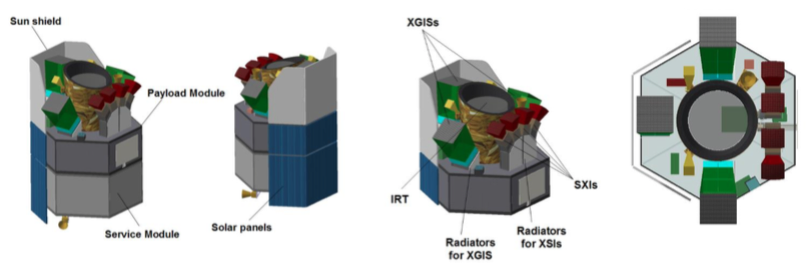}
\caption{The THESEUS baseline configuration.}
\label{f:theseus_baseline}
\end{figure*}

The THESEUS (\emph{Transient High Energy Sky and Early Universe Surveyor}) mission (Figure~\ref{f:theseus_baseline}), proposed for the M5 launch slot in the framework of the ESA Cosmic Vision program, is designed to vastly increase the discovery space of the high energy transient
phenomena over the entirety of cosmic history. 
This is achieved through a payload which combines a wide field sky monitor covering a broad energy band with a soft X-ray instrument with focusing capabilities and a wide field of view, and with an onboard near-IR telescope for immediate transient identification and redshift determination.

The fulfillment of THESEUS scientific goals require the detection of many tens of GRBs from the first billion years (about
30-80) of the Universe (i.e., at redshifts $z > 6$), along the 3 year prime mission lifetime. 
The detection of such a large number of high-$z$ GRBs is obtained thanks to both an increase of about one order of magnitude in sensitivity and an extension of the detector passband down to the soft X-rays (0.3--1 keV), over a broad field of view ($\sim$1 sr) with a
source location accuracy below 1$'$, in order to allow efficient counterpart detection, on-board spectroscopy and redshift measurement, 
and optical and IR follow-up observations. 

To accomplish these goals, the payload includes a set of four monitors based on the \emph{Lobster-Eye} telescope technology (Soft X-ray Imager, SXI), capable of focusing soft X-rays in the 0.3--6 keV energy band over a large FOV ($\sim$1 sr). An onboard infrared telescope (InfraRed Telescope, IRT, sensitive in the 0.7--1.8 $\mu$m band) is also needed, together with spacecraft fast slewing capability (e.g., repointing velocities of about 5$^\circ$--10$^\circ$/min), in order to provide prompt identification of the GRB optical/IR counterpart, on-board redshift determination and spectroscopy of the counterpart and of the host galaxy.

Finally, the payload includes also a broad field of view hard X-ray detection system, the XGIS, whose FOV is larger than that of the SXI, and extending the energy band from 2 keV up to 20 MeV, increasing significantly the capabilities of the mission. 

As the SXI telescopes can be triggered by several classes of transient phenomena (e.g., flare stars, X-ray bursts, etc), the hard X-ray detection system provides an efficient means to identify nearby GRBs and detect other transient sources (e.g., Soft Gamma-ray Repeaters, SGRs) peaking in this energy range. 

The mission profile includes also an onboard data handling unit (DHU) capable of identifying and localizing detected transient events in the SXI and XGIS FOVs, and also the capability of transmitting to ground the trigger time and position of GRBs (and other transients of interest) a few minutes after the detection. 

The baseline mission configuration foresees a launch with Vega-C to a low inclination low Earth orbit (LEO, $\sim$600 km, $<$5$^\circ$), which has the unique advantage of granting a low and stable background level in the high-energy instrument, allowing the exploitation of the Earth's magnetic field for spacecraft fast slewing and facilitating the prompt transmission of transient triggers and positions to the ground.

\section{The XGIS architecture}\label{s:xgis_arch}

\begin{figure}[htbp]
\centering
\includegraphics[width=0.5\textwidth]{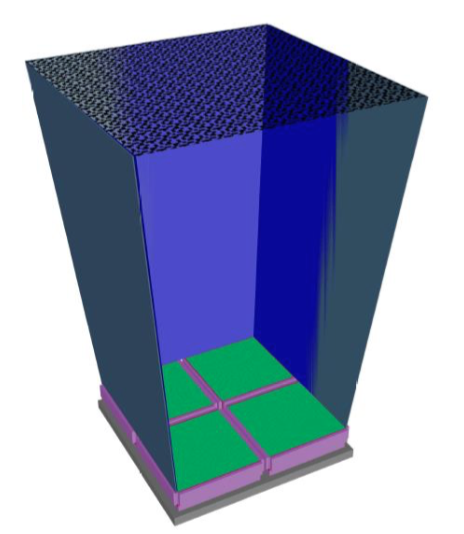}
\caption{One XGIS unit.}
\label{f:xgis}
\end{figure}

The X and Gamma-ray Imaging Spectrometer comprises 3 units, pointed at offset directions in such a way that their FOV partially overlap. Each unit 
(Figure~\ref{f:xgis}) 
has imaging capabilities in the low energy band (2--30 keV) thanks to the presence of a coded mask placed above a position sensitive detector. 
A passive shield around the mechanical structure between the mask and the detector plane will determine the FOV of the XGIS unit for X-rays up to about 150 keV. 
The detector energy range is extended up to several MeVs without imaging capabilities. 
The overall physical dimensions of each XGIS unit are of about 50$\times$50$\times$85~cm$^3$, 
with a total mass of about 37~kg and a power consumption of $\sim$30~W.
The main performance of a single XGIS unit are reported in Table \ref{t:xgis_unit}, as a function of the energy passband.

\begin{table*}[htp]
\caption{Characteristics for a XGIS unit, as a function of the energy}\label{t:xgis_unit}
\begin{center}
\begin{tabular}{lccc}
\hline
& 2--30 keV & 30--150 keV & $>$150 keV \\
\hline
Fully coded FoV & $9^\circ\times9^\circ$ & --- & --- \\
Half sens. FoV & $50^\circ\times50^\circ$ & $50^\circ\times50^\circ$ & --- \\
Total FoV & $64^\circ\times64^\circ$ & $85^\circ\times85^\circ$ (FWZR) & $\sim$2$\pi$ sr \\
Ang. resolution & 25$'$ &  --- & --- \\
Source location accuracy & 5$'$ ($>$5$\sigma$) &  --- & --- \\
Energy resolution (FWHM) & 200 eV at 6 keV & 18\% at 60 keV & 6\% at 500 keV \\
Time resolution & 1 $\mu$s & 1 $\mu$s & 1 $\mu$s \\
On-axis area & 512 cm$^2$ & 1024 cm$^2$ &  1024 cm$^2$ \\
\hline
\end{tabular}
\end{center}
\end{table*}%

\subsection{The dual-readout principle}

\begin{figure}[htbp]
\centering
\includegraphics[width=0.5\textwidth]{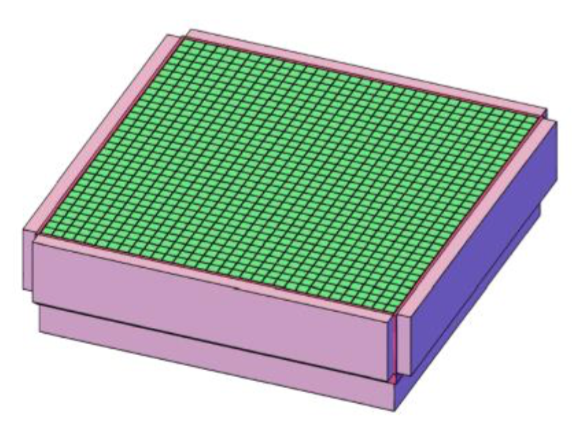}
\caption{One detection module for the sensitive plane of a XGIS unit. Four such modules compose the focal plane.}
\label{f:xgis_detail}
\end{figure}

The detection plane of each unit is made of 4 detector modules, each about 195$\times$195$\times$50 mm$^3$ in size (Figure~\ref{f:xgis_detail}).
Every module includes an array of scintillator bars, read out on top and bottom sides by SDDs, and laterally surrounded by the front-end electronics.

Aiming at designing a compact instrument with a very wide sensitivity band, the XGIS is based on the so-called ``siswich'' concept \citep{marisaldi04,marisaldi05}, exploiting the optical coupling of Silicon detectors with inorganic scintillator bars.

\begin{figure}[htbp]
\centering
\includegraphics[width=0.5\textwidth]{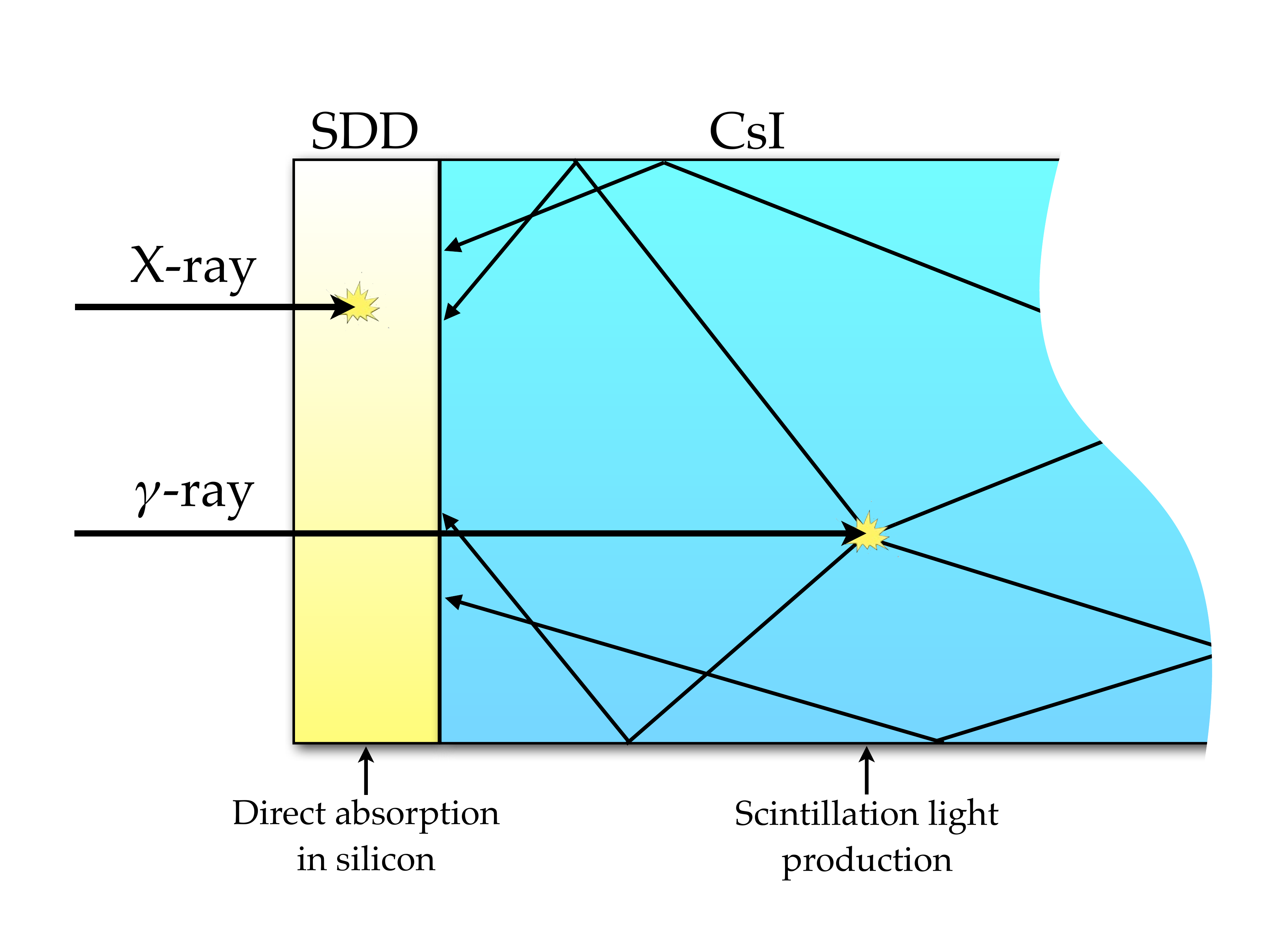}
\caption{The ``siswich'' read-out concept. The left side of the figure represents the ``topmost'' SDD (i.e., the one placed towards the pointing direction).}
\label{f:siswich_concept}
\end{figure}

In this concept (Figure~\ref{f:siswich_concept}), the Silicon detectors on the ``top'' side of the bars play the double role of read-out devices for the signal from the scintillator and of an independent X-ray detector. 
The inorganic scintillator bar, made of thallium-doped caesium iodide, CsI(Tl), is optically coupled at both ends with a Silicon Drift Detector \citep{gatti84}.
Low energy X-rays are stopped in the top-side Silicon detector, while higher energy X-rays and $\gamma$-rays are absorbed in the crystal and the optical scintillation photons are collected by the two SDDs. 
The two types of events are distinguished by pulse shape discrimination techniques, through the different risetimes of the corresponding preamplifier signal.
In the case of ``X-events'', the risetime is dominated by the anode collection time ($\sim$100 ns), while for ``S-events'' the signal rises following the convolution of the characteristic CsI(Tl) scintillation de-excitation time constants (0.68 $\mu$s for 64\% of the scintillation photons, and 3.34 $\mu$s for the remainder) and different light paths, amounting to an effective rise time of a few $\mu$s.

The sensitivity band of the instrument extends from $\sim$1 to $\sim$50 keV (X-events), thanks to the top-side SDD, and from $\sim$20 keV to $\sim$20 MeVs (S-events), thanks to the scintillator, allowing for a partial overlap of the two energy bands.

\subsection{Laboratory characterization of preliminary prototypes}\label{s:prototypes}
\subsubsection{Aim and front-end electronics}
In order to demonstrate the validity of the proposed instrument architecture for the detector plane, a prototype (XGS) is being developed in the framework of an INAF-funded project, and it is currently in its characterisation phase.

\begin{figure}[htbp]
\centering
\includegraphics[width=0.5\textwidth]{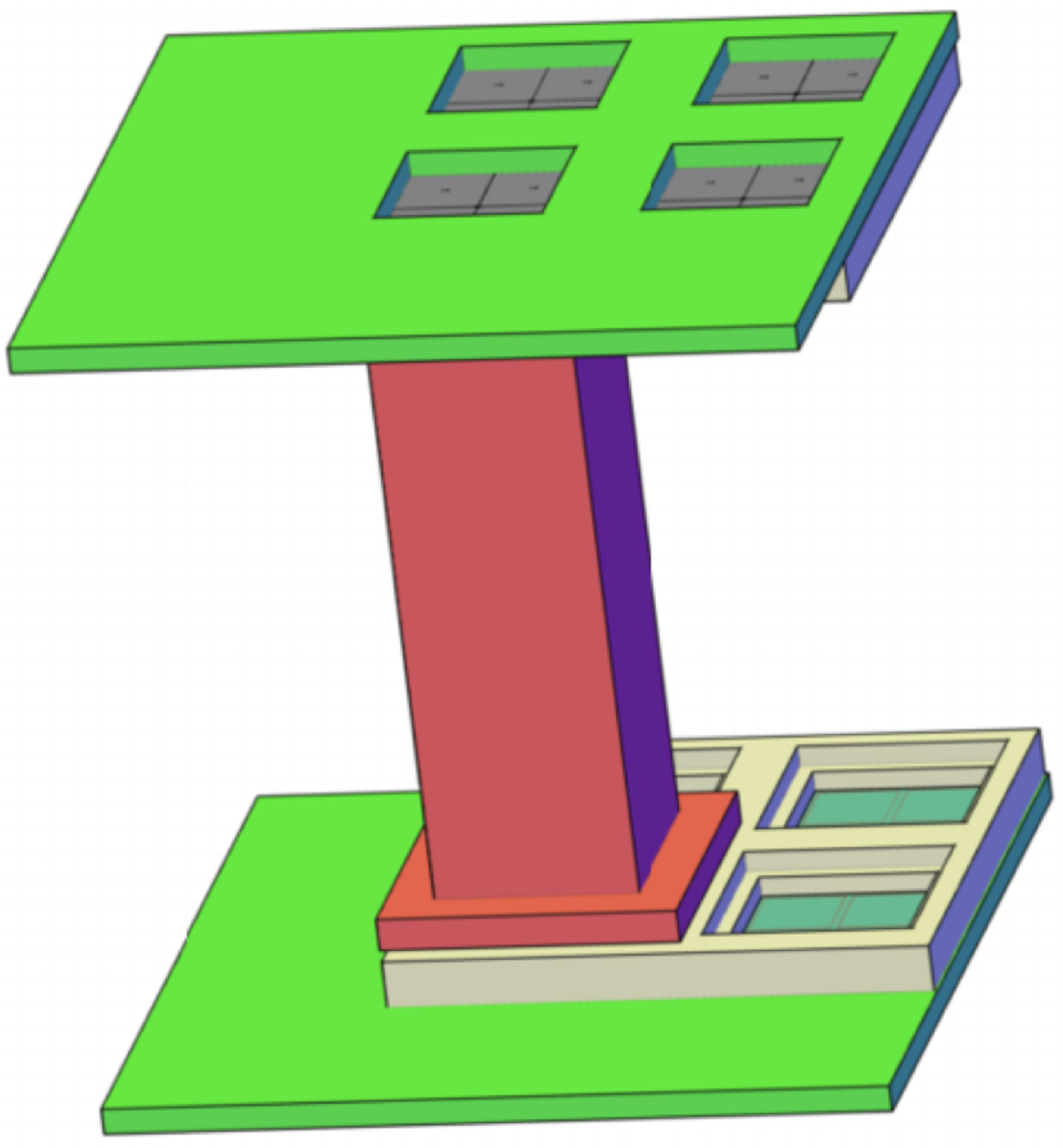}
\caption{Sketch of one module for the architecture demonstrator currently in development phase. A module comprises four SDD matrices, for a total of 16 SDD channels and 16 scintillator bars.}
\label{f:proto_setup}
\end{figure}

\begin{figure}[htbp]
\centering
\includegraphics[width=0.5\textwidth]{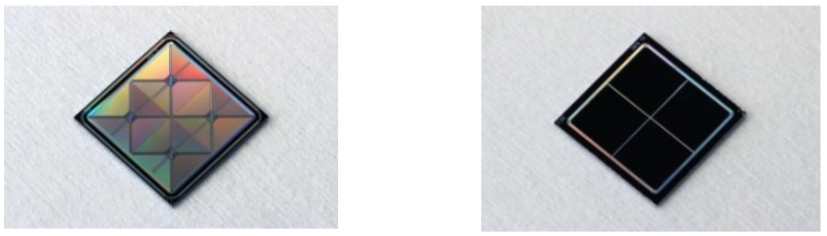}
\caption{The 2$\times$2 SDD array, composed of four 5$\times$5 mm$^2$ single cells. The left panel shows the anode side, while the entrance window for the scintillation light is shown in the right panel.}
\label{f:sdd}
\end{figure}

The prototype comprises four modules or ``quadrants'' (Figure~\ref{f:proto_setup}), each hosting 16 CsI scintillator bars, each with dimensions 5$\times$5$\times$50 mm$^3$.
Each side of a group of 4 scintillator bars is coupled with a 2$\times$2 single-cell SDD array (Figure~\ref{f:sdd}), developed by INFN-Trieste and Fondazione Bruno Kessler (FBK, Trento) within the framework of the ReDSoX collaboration\footnote{\url{http://redsox.iasfbo.inaf.it}},
housed with three other such devices in a front-end electronics board, for a total of 16 channels for each board, 32 for each quadrant. 

\begin{figure}[htbp]
\centering
\includegraphics[width=0.5\textwidth]{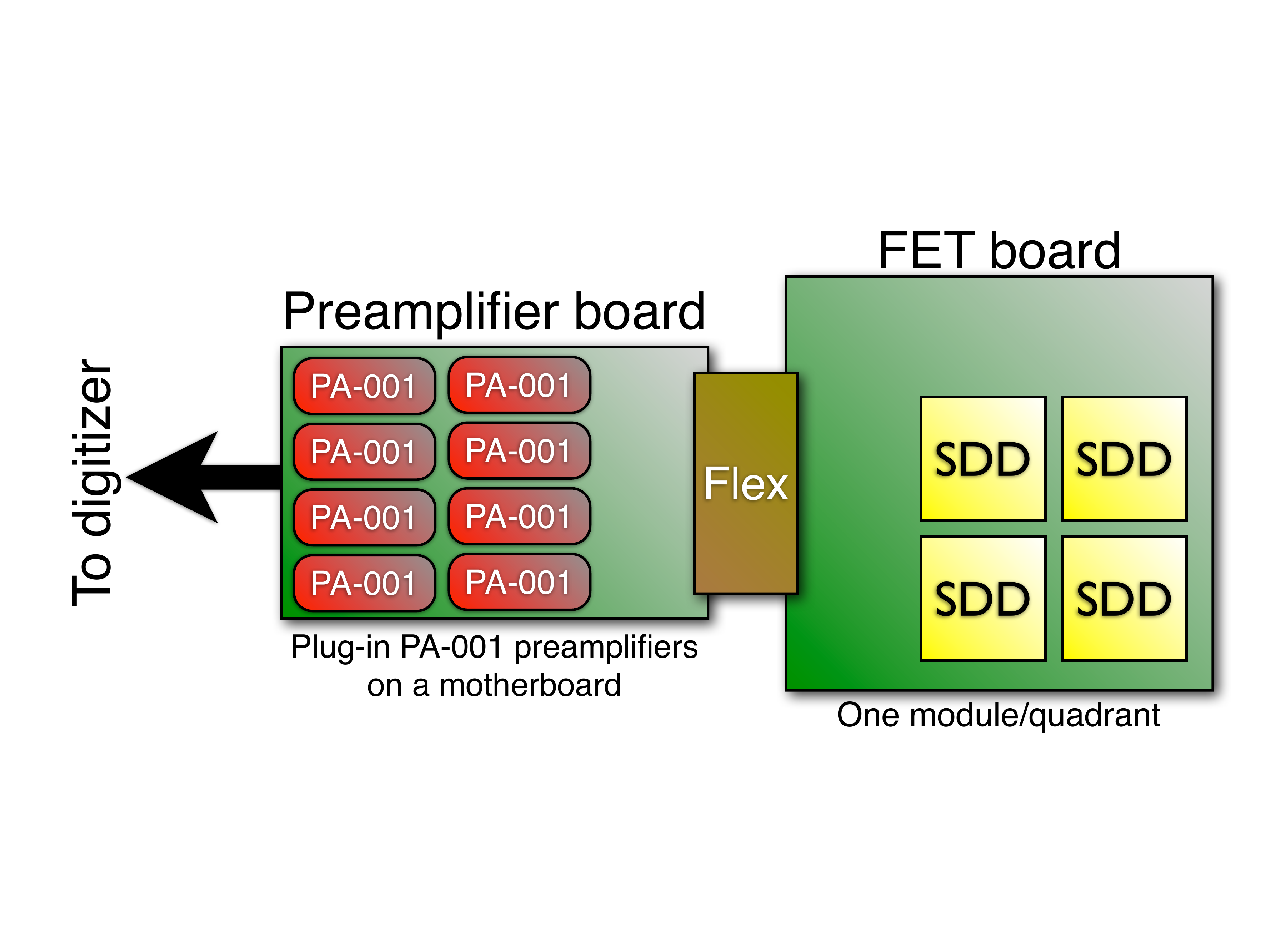}
\caption{The overall front-end setup for the XGS prototype. Four 4-cell SDD matrices (16 channels) are housed in a printed circuit board together with the first-stage preamplifier input FET, and coupled with one end of 16 CsI scintillator bars. The FET board is then connected to a motherboard housing the plug-in PA-001 preamplifiers (8 are shown in this conceptual sketch). The preamplifier signal waveform is therefore sent to an external fast digitizer and then digitally processed by an acquisition computer.}
\label{f:proto_setup2}
\end{figure}

The overall setup is shown in Figure~\ref{f:proto_setup2}.
The front-end electronics is composed of a discrete-element charge sensitive preamplifier (PA-001, developed at INAF/IASF-Bologna), with continuous reset of the feedback capacitor. In order to minimize the stray capacitance, the preamplifier first stage input FET is bonded near the anode, in a dedicated PCB board that distributes also the detector bias voltages.
A flex cable carries the FET output signal to a plug-in motherboard, where single-channel preamplifiers are connected.
The preamplifier output is then fed to a fast digitizer and then digitally processed.

The choice of a digital signal processing (DSP) for the XGS prototype was justified by its large flexibility in this development phase, since it allows simultaneous testing of different shaping algorithms, characterizing the noise performance of the system in a rapid and extremely flexible way.
Of course a custom-designed analog ASIC will be mandatory for an actual space-ready instrument, and the present prototype will allow to drive the design of this front-end electronics component.
While a complete discussion of the results has been partly presented elsewhere \citep{campana16} and will be the subject of a forthcoming paper, in the following some results will be shown.

\subsubsection{Preliminary results}
Several studies have been performed on this module and on a series of preliminary prototypes, aiming at: a) testing the digital processing algorithms; b) investigating the optical coupling between the scintillator bar and the SDDs, and deriving realistic performance figures for the final XGS prototype using a single-bar, 2-channel sub-unit; c) investigating the suitability of the proposed setup for the back-end digital electronics.

\begin{figure}[htbp]
\centering
\includegraphics[width=0.5\textwidth]{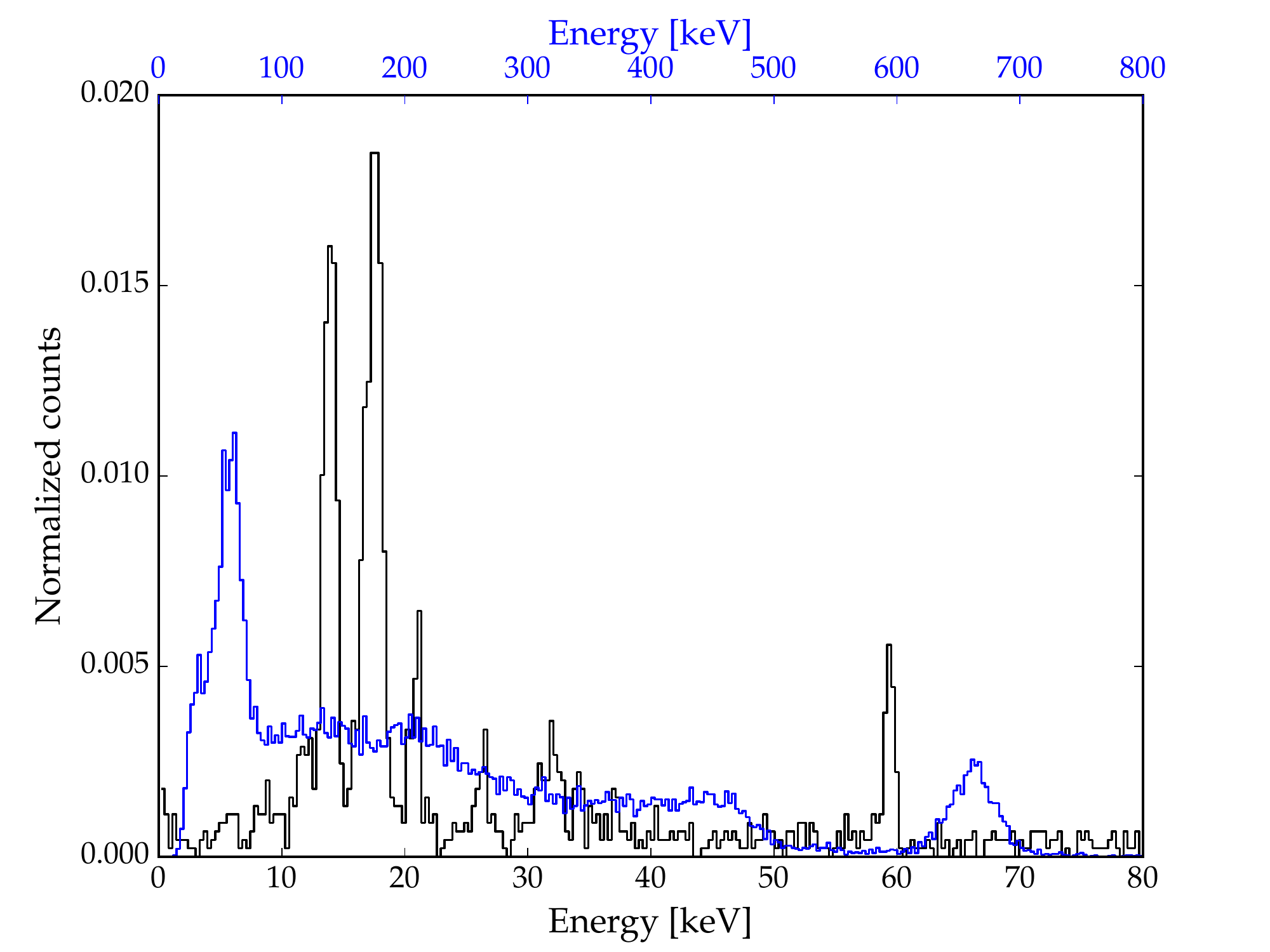}
\caption{Simultaneous acquisition of a $^{137}$Cs plus an $^{241}$Am radioactive source. The black curve (lower $x$-axis) is the spectrum of X-events, while the blue one (upper $x$-axis) is the spectrum of S-events. Both types of events were acquired with a digitizer, identified with pulse shape discrimination and shaped with a trapezoidal filter.}
\label{f:spectrum}
\end{figure}

In Figure~\ref{f:spectrum} it is shown a simultaneous acquisition at room temperature of two radioactive sources, $^{137}$Cs (line at 662 keV plus Compton continuum) and $^{241}$Am (low energy X-ray lines between 11 and 26~keV, and line at 59~keV). 
Both types of events were acquired simultaneously by the digitizer and then distinguished with pulse shape discrimination on the rising edge. 
Optimized digital filters were then applied separately for the two categories.
The good performance of the digital signal processing methods are thus demonstrated, showing a comparable result with respect to a more traditional acquisition chain composed by an analog shaping amplifier coupled to a commercial multichannel analyzer.

\begin{figure}[htbp]
\centering
\includegraphics[width=0.49\textwidth]{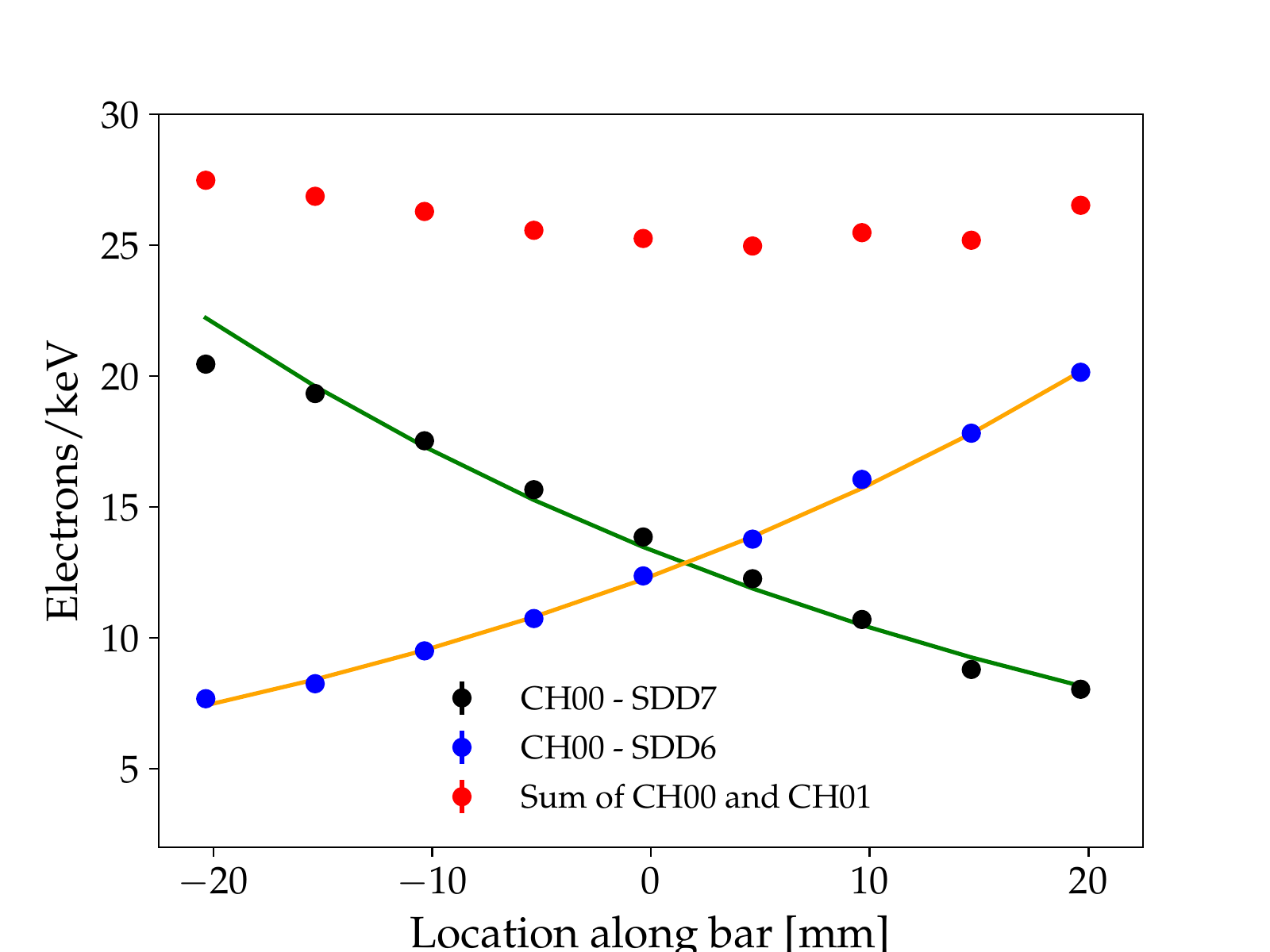}
\includegraphics[width=0.49\textwidth]{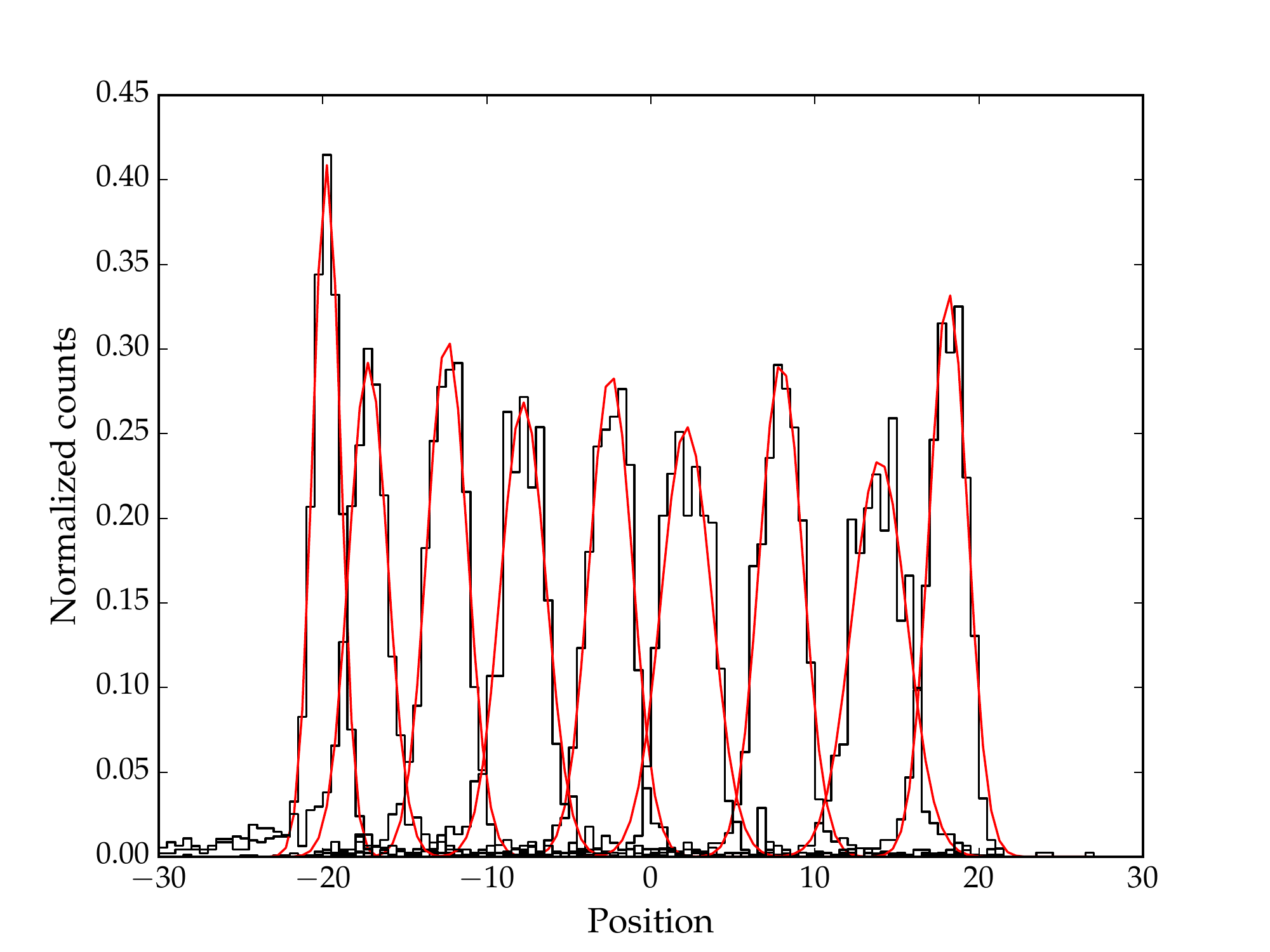}
\caption{\emph{Upper panel:} Light attenuation along the bar. The signal from both detectors is shown, together with a best-fit exponential attenuation line. \emph{Lower panel:} Reconstructed position for various measurements along the bar.}
\label{f:protoxgs2_results}
\end{figure}

\begin{figure}[htbp]
\centering
\includegraphics[width=0.5\textwidth]{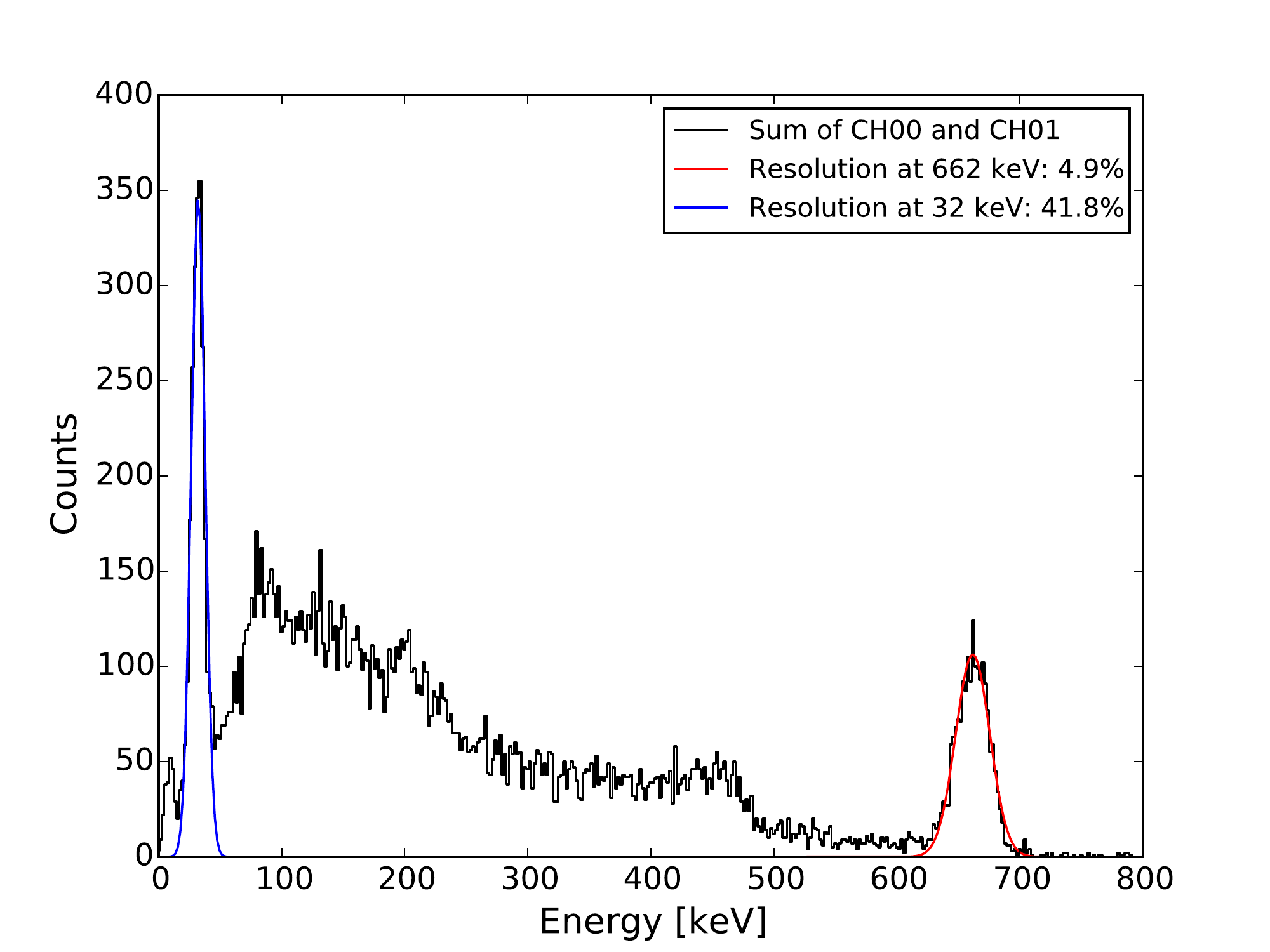}
\caption{Spectrum of a collimated $^{137}$Cs radioactive source, placed in the middle of the scintillator bar, acquired summing the scintillation light signal collected by the SDDs at both ends. Gaussian fits of the 662 keV and 32 keV peaks are also shown.}
\label{f:protoxgs2_spectrum}
\end{figure}

In Figure~\ref{f:protoxgs2_results}, upper panel, the scintillation light attenuation measured by the two detectors at both ends of a scintillator bar is shown. The attenuation coefficient is $\sim$0.021 cm$^{-1}$, and the scintillation signal collected by the two photo-detectors (dependent on the light yield of the scintillator crystal, the optical coupling between crystal and SDD, and on the SDD quantum efficiency) is $\sim$27 e$^-$/keV, consistent with previous measurements \citep{marisaldi06, labanti08}.  
In the lower panel of Figure~\ref{f:protoxgs2_results} the position reconstruction along the bar, derived from the different signal collected at the two ends \citep[see][for a discussion on this method]{labanti08}, is shown. The average FWHM of the reconstructed position is about 3~mm. Considering the source spot size in these measurements ($\sim$2 mm), the derived intrinsic resolution in position for $\gamma$-ray events is around 2.2 mm (FWHM).
In Figure~\ref{f:protoxgs2_spectrum} a spectrum obtained with the source at the bar center is shown. The measured resolution at 662~keV is 4.9\%, with a lower threshold of about 20~keV, thus confirming the expected results and the good performance of the system.

\subsection{Imaging with coded mask}

The finely pixellated structure of the XGIS detection modules makes them ideal for the use in connection with a coded mask in order to obtain an instrument with good imaging capabilities over a wide field of view. Imaging will be optimized in the softer energy range thanks to a light-weight mask able to modulate the incoming photons in the   range of the X-events. The mask will become increasingly ineffective at higher energy, where the XGIS will operate as a non-imaging collimated spectrometer. However, given the spectra expected from typical X and $\gamma$-ray sources, the largest number of photons will be obtained at low energy, thus allowing to localize the source with a good precision thanks to the coded mask imaging system.

The baseline design of each unit, that will be optimized during the detailed study phase, is based on a square mask of overall size $\sim$50$\times$50 cm$^2$ placed at a distance of 70~cm above the detectors.  This will give a zero-sensitivity field of view of $\sim$60$\times$60 deg$^2$ and a fully coded field of view of $\sim$10$\times$10 deg$^2$. 
The mask will be made of stainless steel, with a thickness of 500 $\mu$m, and will be based on a pattern which allows self-support
in order to guarantee the maximum transparency of the open elements. 

A stainless steel mechanical structure (thickness 100 $\mu$m) will connect the mask with the detector, supporting 4 Tungsten slats 45 cm high with an optimized variable thickness (0.5--0.3 mm) to act as a passive shield for the imager system (in 1--50 keV energy range) and to define the field of view below $\sim$150 keV. In the latter range, the resulting FOV is $\sim$50$\times$50 deg$^2$ (FWHM) and  $\sim$85$\times$85 deg$^2$ (FWZR). The three units will be combined in such a way that their fields of view will partially overlap: by offsetting two of them by 35$^{\circ}$ at opposite directions the FOV delimiter guarantees an average XGIS effective area of $\sim$1400 cm$^2$ in the SXI FOV (104$\times$31 deg$^2$).

\begin{figure}[htbp]
\centering
\includegraphics[width=0.5\textwidth]{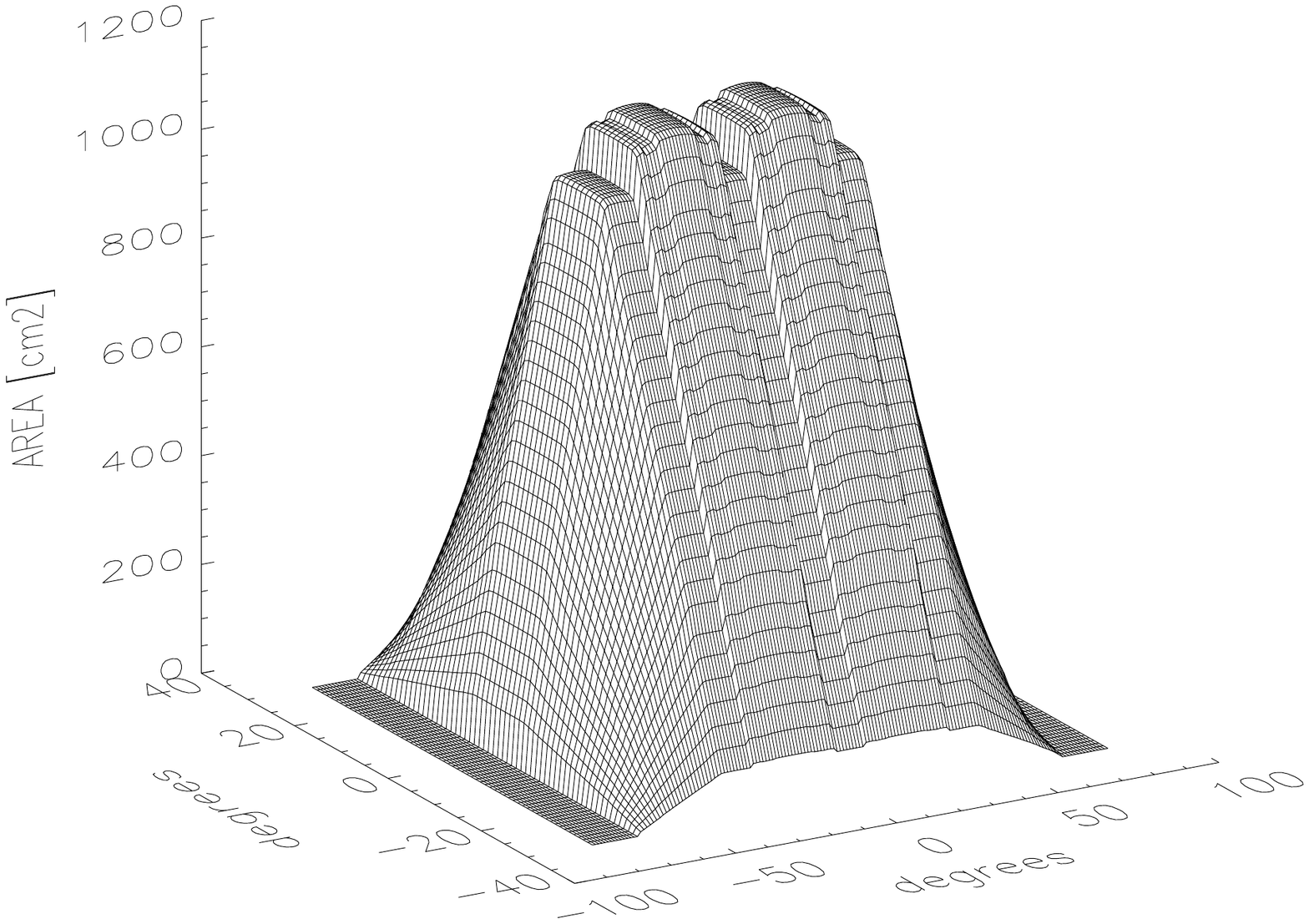}
\caption{Effective area of the three units for offsets of 35$^{\circ}$,  giving a partial overlap of the fields of view of the three units. }
\label{f:fov}
\end{figure}

\begin{figure}[htbp]
\centering
\includegraphics[width=0.5\textwidth]{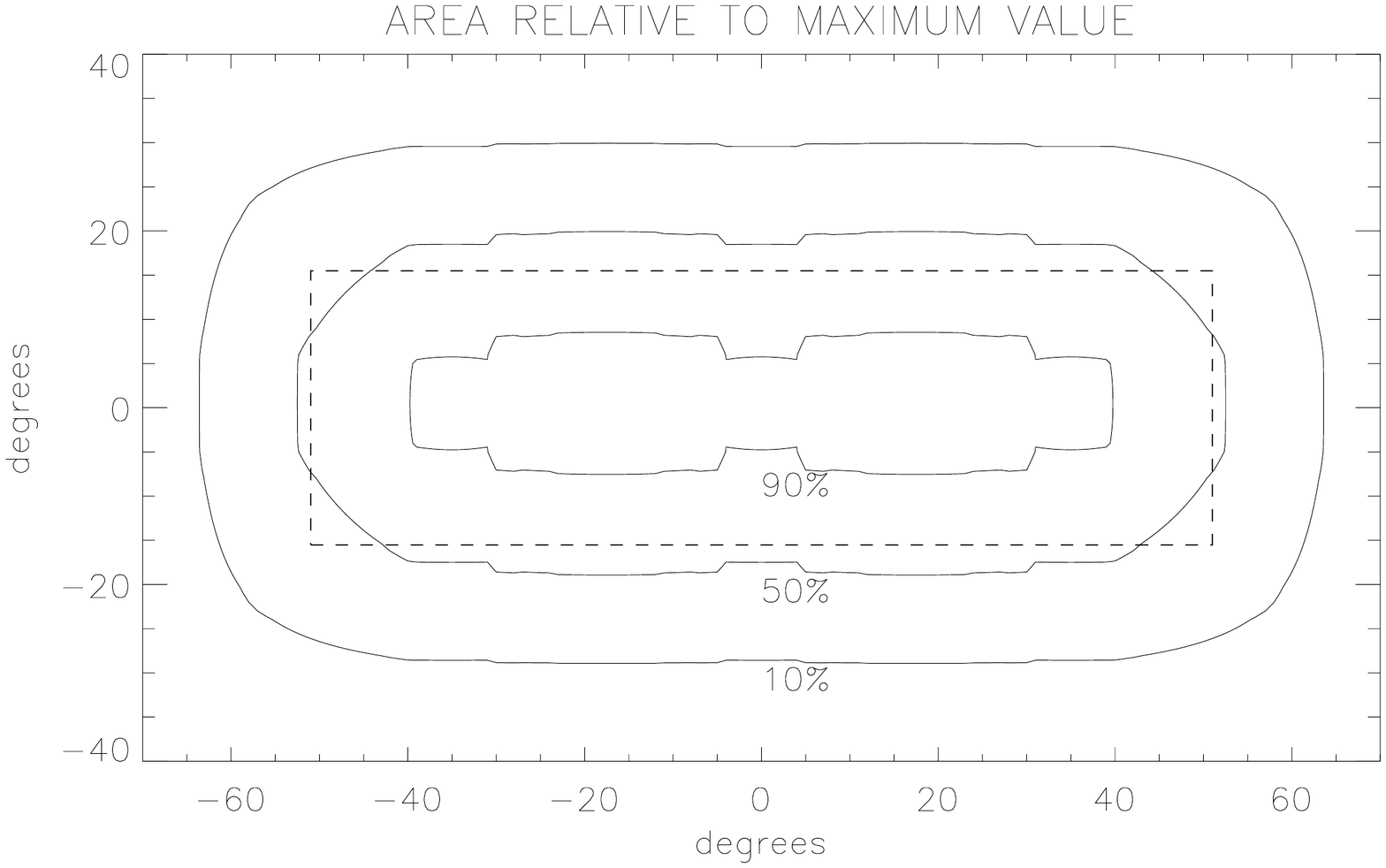}
\caption{XGIS imaging field of view (contours give the effective area compared to the on-axis value) compared with that of the SXI (dashed rectangle).}
\label{f:fov_cont}
\end{figure}

Figure~\ref{f:fov} shows how the effective area of the combined imaging systems of the three units varies in the field of view. 
As can be seen in Figure~\ref{f:fov_cont}, this configuration allows to cover with good imaging sensitivity the field of view of the SXI instrument.

\subsection{The front-end electronics}
The Front End Electronics (FEE) design for XGIS will meet the requirements of minimizing the overall noise readout of a SDD taking into account the limited power budget. For the entire XGIS the overall number of SDDs to be read out is 24576 (32$\times$32 bars per module, four modules in an unit, three units). 
Due to this large number of elements, the technology used in the FEE will be based on ASICs. Certain operations on the signal, as pre-amplification, must be done as close as possible to the SDD to reduce the coupling stray capacitance. 
As an optimum solution, two different ASICs will be realized. Each ASIC-1 will deal with two SDD signals and will perform pre-amplification (and/or shaping) and buffering to transfer the signal to the rest of the processing electronics few tens of cm away, while the ASIC-2 (each chip ideally dealing with 8 or 16 channels) will take care of the trigger generation and  discrimination.
This overall design baseline will lead to 512 ASIC-1 and 128 ASIC-2 per module.
The ASIC-1 and ASIC-2 design will be based on the VEGA-ASIC design developed at Politecnico di Milano and Universit\`a di Pavia for the readout of SDD devices in the REDSOX project frame \citep{ahangarianabhari14}. 
ASIC-1s will be placed on the top and bottom of each XGIS module while ASIC-2s will be placed on the lateral side of each XGIS module. 

The Digital Front End Electronics (DFEE) will interface the FEE with the Instrument Data Handling Unit (IDHU) and the Payload Power Supply Unit (PSU). Each XGIS module will have its DFEE whose main functions are to interface the stretched ASIC-2 output and to provide Analog to Digital Conversion (ADC), event time tagging, detector and module identification.

At XGIS unit level (4 modules) the Unit Control Electronics (UCE) will manage the low voltage (LV, 3.3 V) and medium voltage (MV, 180 V) power supply post-regulation and filtering and distribution, besides the telemetry, housekeeping and tele-command interface and management.

\section{Instrument functions and performance}\label{s:performance}
\subsection{The Data Handling Unit and its functions}
The whole XGIS background data rate (3 units) towards the instrument Data Handling Unit (DHU) is of the order of 6000 events/s in the 2--30 keV range and about 3700 events/s above 30 keV. Each event received by DHU will be identified with a word of 64 bits (4 for module address, 10 for bar address, 10 for signal amplitude of the fast top channel, 11 for signal amplitude of the slow top channel, 11 for signal amplitude of the slow bottom channel, 18 for time).
The DHU functions will include the discrimination between X and S events: fast and slow channels are required for the pulse shape discrimination on the topside SDDs.
For the latter type, the interaction position inside the bar by weighting the signals of the 2 SDD will be evaluated. Combining this information with the address of the bar (5$\times$5 mm in size) each module becomes a 3D position sensitive detector.
In order to provide critical physical information on the background and transient events, the DHU will also continuously calculate the acquisition flow of the detector with a sampling rate of 1~Hz, independently of any on-board transient trigger. These informations will be provided on different energy bands and geometric regions (units, sub-units). 
In the 2--30 keV range and for each unit, the DHU produces images of the FOV in a defined integration time, holding also in a memory buffer all the XGIS data, rates and images of the last 100 (typical) seconds with respect to the current time, in order to allow for an image-based trigger (see Sect.~\ref{s:imagetrigger}).
Maps of the three unit planes will be produced, with event pixel by pixel histograms in different energy bands and with selectable integration times.

\subsection{XGIS and the GRB trigger system}

XGIS will contribute to the THESEUS overall GRB trigger system in different ways: 
\subsubsection{Qualification of the SXI triggers} 
Because of its energy sensitivity, the SXI will produce several triggers on different transient types (not only GRBs). Therefore, the primary role of XGIS is to qualify the SXI triggers as true GRBs. The basic algorithm for GRB validation is based on an evaluation of the significance of the count rate variation: from the SXI direction given to the event, it is identified one of the three XGIS units in which the event has potentially been detected; then, it will be searched an excess of counts in the modules of this unit in the bands 2--30 keV and 30--200 keV with respect to the average count rate continuously calculated by DHU.
\subsubsection{Autonomous XGIS GRB trigger based on data rate} 
The autonomous GRB trigger for XGIS inherits the experience acquired with the Gamma Ray Burst Monitor (GRBM) aboard BeppoSAX and that acquired with the MCAL instrument aboard AGILE (Fuschino et al., 2008) and concerns all the modules of the 3 XGIS units. For each module, the abovementioned energy intervals (2--30 and 30--200 keV) are considered for the trigger. The mean count rate of each module in each of these bands is continuously evaluated on different time scales (e.g., 10 ms, 100 ms, 1 s, 10 s). A trigger condition is satisfied when, in one or both of these energy bands, at least a given fraction (typically $\ge$3) of detection modules sees a simultaneous excess with a significance level of typically 5$\sigma$ on at least one time scale with respect to the mean count rate.
\subsubsection{Autonomous XGIS GRB trigger based on images}\label{s:imagetrigger}
For each XGIS unit, the 2--30~keV actual images will be confronted with reference images derived averaging $n$ (typically 30) previous images, and a spot emerging from the comparison at a significance level of typically 5$\sigma$ will appear. This allows also for the localisation of events outside the SXI FoV. 

If one of the previous trigger conditions is satisfied, event by event data starting from 100 s before the trigger are transmitted to ground.
The duration of this mode lasts until the counting rate becomes consistent with background level.

\subsection{Telemetry requirements}
For the study of transient or persistent sources, different transmission modes will be selected starting from the photon list and the histogram maps of the units. Typically the telemetry load will be maintained below 2 Gbit/orbit transmitting at low energies ($<$30 keV) event histograms in various energy channels (e.g. 32 channels) with variable integration times (e.g. 64 s).
Above the 30 keV the entire photon-by-photon event list will be trasmitted.
In particular observations, such as for example in the case of crowded fields a photon by photon transmission in the total energy range will be selected for a total maximum telemetry load of 3 Gbit/orbit.
Of course, in the case of a GRB trigger all the information available photon by photon is transmitted with a maximum load of 1 Gbit.

\subsection{XGIS sensitivity}
\begin{figure}[htbp]
\centering
\includegraphics[width=0.5\textwidth]{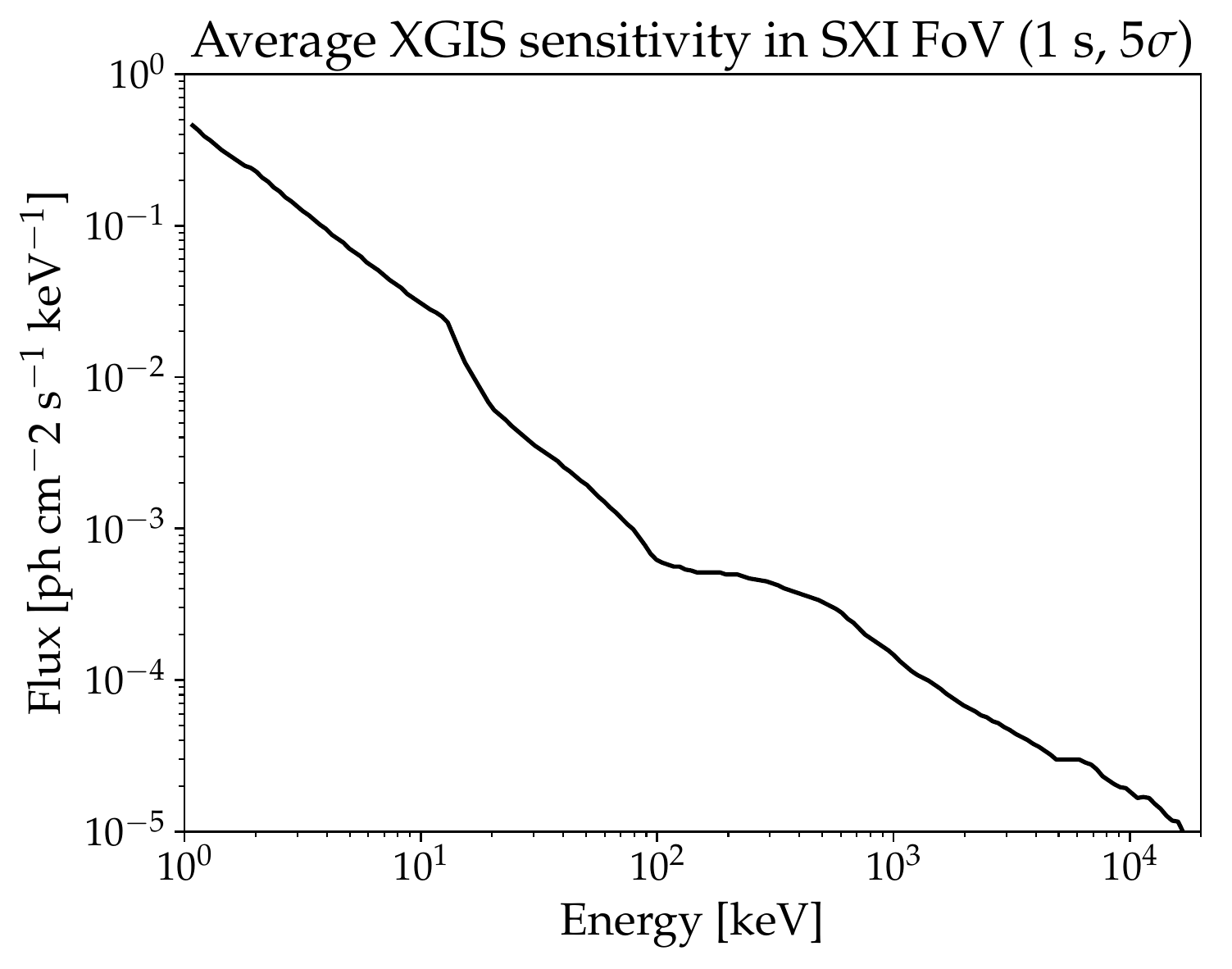}
\caption{XGIS sensitivity for a 1 second exposure (5$\sigma$ level), as a function of the energy.}
\label{f:sens_5s1s}
\end{figure}

\begin{figure}[htbp]
\centering
\includegraphics[width=0.5\textwidth]{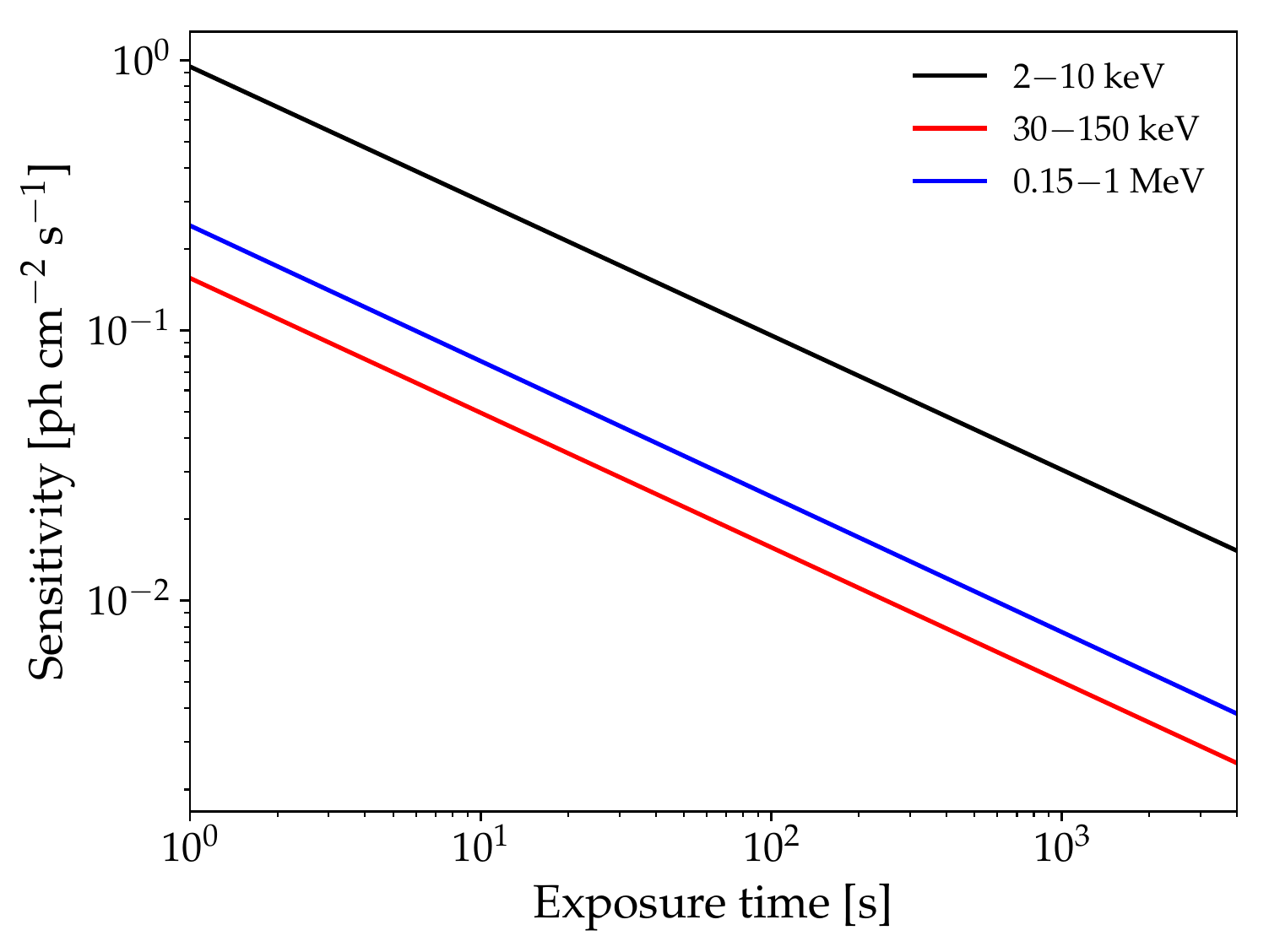}
\caption{XGIS sensitivity in different energy bands, as a function of the exposure.}
\label{f:sens_exposure}
\end{figure}

\begin{figure}[htbp]
\centering
\includegraphics[width=0.5\textwidth]{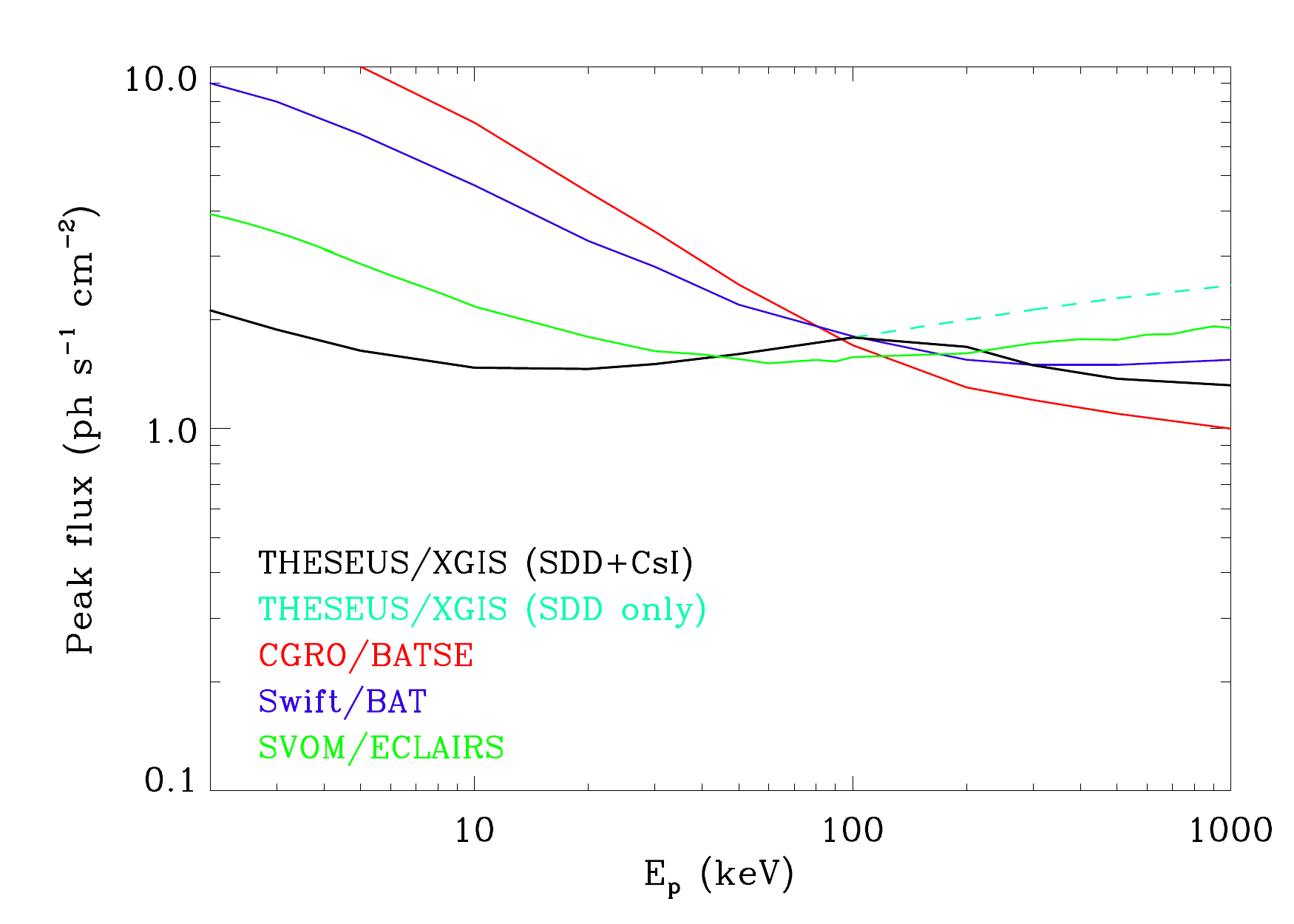}
\caption{Sensitivity of the XGIS to GRBs in terms of minimum detectable photon peak flux in 1s (5$\sigma$) in the 1--1000 keV energy band as a function of the GRB spectral peak energy. A Band function with $\alpha = -1$ and $\beta = -2$ has been assumed \citep{band03}. Swift/BAT and SVOM/ECLAIRs data from \cite{band06} and \cite{godet12}.}
\label{f:xgis_sens_comparison}
\end{figure}

XGIS will be a sensitive instrument throughout a wide energy band.
The 5$\sigma$ XGIS 1 s sensitivity with energy in the SXI FOV, is shown in Figure~\ref{f:sens_5s1s}, as a function of the energy. In Figure~\ref{f:sens_exposure} the XGIS flux sensitivity dependence on the observation time at a significance of 5$\sigma$ in different energy ranges. In Figure~\ref{f:xgis_sens_comparison}, the XGIS sensitivity vs. GRB peak energy is compared with that of other instruments.
As can be seen, the combination of large effective area and unprecedented broad energy band provides a much higher sensitivity w/r to previous (e.g., CGRO/BATSE), present (e.g., Swift/BAT) and future (e.g., SVOM/ECLAIRs) in the soft energy range, while keeping a very good sensitivity up to the MeV range.

\section{Summary and conclusions}\label{s:conclusions}

A sensitive and broad-band X and $\gamma$-ray instrument is needed to fulfill the THESEUS scientific objectives, in particular
to reliably identify GRBs and their counterparts with autonomous trigger and trigger validation capabilities. Moreover, the proposed XGIS design will allow also to measure high-energy transients on short timescales with also spectroscopic sensitivity, allowing to investigate in detail the energetic and temporal evolution of these sources.
Of course, several trade-offs and improvements on the basic design outlined in this paper can be foreseen. The field of view of a single unit, the coded mask open fraction (and its impact on efficiency and sensitivity) but also the choice of scintillator thickness and type (and its impact on efficiency and spectral resoluton), besides the detailed triggering logic, can and should be optimized in a detailed future assessment phase in order to maximize THESEUS scientific returns.

\bibliography{bibliography}
\bibliographystyle{aa}

\end{document}